\documentclass[fleqn,twoside]{article}
\usepackage{espcrc2}

\usepackage{graphicx}\usepackage{epsfig}
\usepackage[figuresright]{rotating}

\newcommand{\AmS}{{\protect\the\textfont2
  A\kern-.1667em\lower.5ex\hbox{M}\kern-.125emS}}

\title{Rare Decays as a Probe for New Physics}

\author{T.~Hurth\address{CERN, Dept.\ of Physics, Theory Unit, CH-1211 Geneva 23, Switzerland\\ SLAC, Stanford University, Stanford, CA 94309, USA}
\thanks{\small  Heisenberg Fellow}}

\begin{document}

\begin{abstract}

We discuss the indirect search for new degrees of freedom 
beyond the standard model,  within flavour physics. 
In particular, we analyse the minimal flavour violation  hypothesis 
and its phenomenological implications, especially the large-$\tan \beta$ 
scenario  in supersymmetric models, and also compare it  with the 
constrained minimal flavour violation scenario. 
Moreover, we briefly discuss  some recent  progress in inclusive $b \to s$ transitions and   present a status report of   the so-called  
$K \pi$ \mbox{puzzle}. \hspace{3.6cm}    {\it CERN-PH-TH/2006-267,SLAC-PUB-12267}
\vspace{1pc}
\end{abstract}

\maketitle

\section{INTRODUCTION}
With the running $B$-factory experiments Babar  at SLAC  and Belle at KEK,
the forthcoming  $B$-physics programmes of the LHC experiments, especially LHCb, at CERN  and the future options of $B$ experiments at Super-$B$ 
factories  and  at future linear colliders, 
$B$ \mbox{physics} is one important focus in particle physics today.

It is well known that rare $B$ decays,  as  flavour-changing 
neutral currents (FCNCs), are particularly sensitive to new physics.
The present data from the $B$-physics experiments  
already imply     significant restrictions for the parameter 
space of new physics models  
and lead  to important 
clues for the direct search for new particles and the model building 
beyond the standard model (SM).  
After new physics will have  been discovered, especially when the mass 
scale of the new physics will have been  fixed, such observables,   
and also correlations between collider and flavour 
observables, will play an important role in analyzing the 
underlying new dynamics.

In particular, there is a flavour problem to solve in any viable new physics model,
namely why FCNCs  are suppressed.
In supersymmetry the flavour problem is directly linked to the 
crucial question of how supersymmetry is broken. 
Moreover, there is  a corresponding CP problem: while the CKM prescripton 
has passed its first precision tests, the problem arise of finding the 
mechanism by which the often numerous additional CP phases are suppressed in a 
new physics model. For example, 
there are very   stringent bounds on the 44  phases of the 
minimal supersymmetric standard model (MSSM). 

One of the main difficulties in examining the observables in 
$B$  physics is the influence of the strong interaction.
If new physics does not show up in flavour  physics 
through large deviations, as recent experimental data indicate, 
one has to focus 
          on theoretically clean variables such as 
inclusive rare $B$ decays,  which are dominated by perturbative contributions 
or specific ratios of exclusive modes such as  CP or charge 
asymmetries.
It is important to calculate those specifically suitable observables 
to very high precision in order to exploit their sensitivity to possible 
degrees of freedom beyond the SM.

 In the indirect search for new physics, it is also mandatory  to go 
 beyond  the analysis of branching ratios and to measure 
more complex kinematical distributions such as  CP, forward-backward, and isospin 
asymmetries to detect subtle patterns and to distinguish between the 
various scenarios beyond the SM.

The paper is organized as follows. In the next section we discuss 
the minimal flavour violation hypothesis and its phenomenological implications.
Moreover, two analyses of the large $\tan \beta$ scenario in 
supersymmetric models  are briefly  reviewed. In section 3
we briefly discuss  some  specific opportunities 
for the new-physics search within
$b \to s$ transitions and give more details  on the present status 
of the so-called $K \pi$ \mbox{puzzle.} The reader will find 
a detailed discussion of the  $b \to s$ transitions 
in a forthcoming review~\cite{hurthnew}.

\section{MINIMAL FLAVOUR VIOLATION AND BEYOND}
There are two general  approaches to new physics, which are most suitable 
in the present situation where no direct evidence for new degrees 
of freedom beyond the SM exists. 

\begin{table*}
\caption{\sf Bounds on rare decays in constrained MFV \label{brMFV}}
\vspace{0.5cm}
\begin{center}
\begin{tabular}{|c|c|c|c|c|}
\hline
{Branching Ratios} &  MFV (95\%) &  SM (68\%) &  SM (95\%) & exp
 \\ \hline
${\cal B}(K^+\to\pi^+\nu\bar\nu)\times 10^{11}$ & $< 11.9$ & $8.3 \pm 1.2$ &  $[6.1,10.9]$
& $(14.7^{+13.0}_{-8.9})$ 
\\ \hline
${\cal B}(K^0_{L}\to\pi^0\nu\bar\nu)\times 10^{11}$  & $< 4.59$ &  $3.08 \pm 0.56$ &  $[2.03,4.26]$ &
 $ < 5.9 \cdot10^{4}$  
\\ \hline
${\cal B}(K^0_{L} \to \mu^+ \mu^-)  \times 10^{9} $ & $< 1.36$ & $0.87 \pm 0.13$ &  $[0.63,1.15]$ & -
\\ \hline
${\cal B}(B\to X_s\nu\bar\nu)\times 10^{5}$ & $<5.17$ &  $3.66 \pm 0.21$ &  $[3.25,4.09]$
&  $<64 $ 
\\ \hline
${\cal B}(B\to X_d\nu\bar\nu)\times 10^{6}$ &  $<2.17$ & $1.50 \pm 0.19$ &  $[1.12,1.91]$
& -
\\ \hline
${\cal B}(B^0_s\to \mu^+\mu^-)\times 10^{9}$ &  $< 7.42$ & $3.67 \pm 1.01$ &  $[1.91,5.91]$
& $<2.7\cdot 10^{2}$  
\\ \hline
${\cal B}(B^0 \to \mu^+\mu^-)\times 10^{10}$ &  $< 2.20$ & $1.04 \pm 0.34$ &  $[0.47,1.81]$
& $<1.5 \cdot 10^3$ 
\\ \hline
\end{tabular}
\end{center}
\end{table*}

While a model-independent analysis takes into account the 
possibility of new flavour structures, which  are parametrized by  
model-independent parameters, an analysis within the 
minimal flavour violation (MFV) hypothesis assumes that the flavour and the CP 
symmetry are  broken as in the SM;  it 
essentially requires that all flavour- and CP-violating interactions be 
linked to the known structure of Yukawa couplings 
(called $Y_U$ and $Y_D$ in the following).  
A renormalization-group-invariant definition of MFV 
based on a symmetry principle 
is given in~\cite{Chivukula:1987py,Hall:1990ac,D'Ambrosio:2002ex}; 
this  is mandatory for  a consistent  
effective field theoretical analysis
of new physics effects. In fact, a low-energy effective theory 
with all SM fields including one or two Higgs doublets is constructed; 
as the only source of $U(3)^5$ flavour symmetry breaking, 
the ordinary Yukawa couplings are introduced as background values of fields  
transforming  under the flavour group (`spurions')~\cite{D'Ambrosio:2002ex}.

\begin{figure}[t]
\begin{center}
\includegraphics[width=7.3cm,angle=-90]{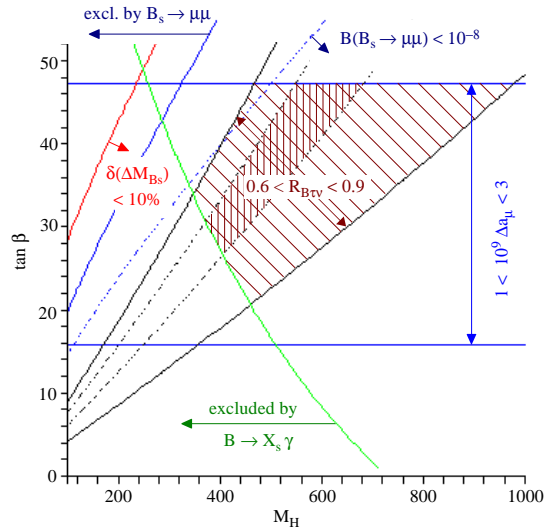}
\vspace{-1cm}
\caption{\sf Constraints on the charged-Higgs mass -- $\tan \beta$ plane in 
the MFV within the  MSSM. \label{Gino}} 
\end{center} 
\end{figure}

In the construction of the effective field theory,  operators with arbitrary 
powers of the dimensionaless $Y_{U/D}$ have to be considered in principle. 
However, the specific  structure of the SM,  with  its hierarchy 
of CKM matrix  elements and quark masses,  drastically reduces 
the number  of numerically  relevant operators. For example, it can be shown  
that in MFV models with one Higgs doublet, all FCNC  processes with  external 
$d$-type quarks are governed by 
the following combination of spurions due to the dominance of the top 
Yukawa coupling $y_t$:
\begin{equation}
(Y_U Y_U^\dagger)_{ij} \approx y_t^2  V^*_{3i} V_{3j}\,,  
\end{equation} 
where  a basis is used in which the  $d$-type quark Yukawa is diagonal.

There are two strict predictions in this general 
class of models, which have to be  tested. First, the MFV hypothesis implies 
the 
usual CKM relations between $b \to s$, $b \to d$, 
and  $s \to d$ transitions. For example, this relation allows 
for upper bounds  on new-physics effects in 
${\cal B}(\bar B \to X_d\gamma)$, and ${\cal B}(\bar B \to X_s \nu\bar \nu)$ using experimental data or bounds from ${\cal B}(\bar B  \to X_s\gamma)$, and 
${\cal B}(K \to \pi^+ \nu\bar \nu)$ respectively. 
This emphasizes the need for 
high-precision measurements of $b \to  s/d$ , but also of 
$s \to  d$ transitions such as  the rare kaon decay 
$K \to \pi \nu\bar\nu$.

The second prediction is  that  the CKM phase is the only                            source of CP violation. This implies that any phase 
measurement as  in 
$B \to \phi K_s$ or $\Delta M_{B_{(s/d)}}$ is  not sensitive 
to  new physics. Note that there is also a
RG-invariant extension of the MFV concept allowing for flavour-blind phases 
\cite{Hurth:2003dk}; these  lead to non-trivial CP  effects,  
which get, however, strongly constrained by flavour-diagonal observables 
such as  
electric dipole moments (see for an example \cite{Hurth:2003dk}).

The usefulness of MFV-bounds/relations is obvious; any 
measurement beyond those bounds \mbox{indicate} the existence of new 
flavour structures.

In Ref.~\cite{Bobeth:2005ck}, upper bounds 
for rare decays in {\it constrained} MFV (CMFV) models 
at $95 \%$ probability are presented, see Table \ref{brMFV}. 
Furthermore, 
an upper bound on the $B_q -\bar B_q$ ($q=d,s$) mixing
in CMFV was  established more recently~\cite{Blanke:2006yh}.
However, there is a subtlety involved. Those bounds 
are based on a constrained  version of MFV 
where the {\it additional} 
dynamical assumptions are  used  
that the relevant operators in the electroweak 
effective Hamiltonian 
for weak decays are the same as in the SM and that new physics 
            only leads to changes of the effective couplings, the Wilson 
coefficients, but not of the CKM factors of those operators. 
Clearly, those assumptions allow for  additional  relations 
between different $B$ and kaon observables (see also \cite{BurasII}), 
but their violation do not necessarily indicate the existence of new flavour 
structures. Moreover, they  explicitly rule out -- as 
most important difference to the general  MFV concept --  
new scalar operators and, thus, large 
$\tan \beta$ effects, which are not necessarily based 
on new flavour structures.

It is well known that scenarios including  two Higgs doublets with large 
$\tan \beta = O(m_t/m_b)$  allow for  the unification of top and bottom 
Yukawa couplings, as predicted in grand-unified models 
(see \cite{Hall:1993gn}), and 
for sizable new effects in helicity-suppressed decay modes (see 
\cite{Hamzaoui:1998nu,Babu:1999hn}).
There are more general MFV relations existing  in this scenario due 
to the dominant role of scalar operators. However, since  
$\tan \beta$ is large,  
there is a new combination of spurions numerically relevant in 
the construction of higher-order MFV effective \mbox{operators,} namely

\begin{equation}
(Y_D Y_D^\dagger)_{ij} \approx y_d^2  \delta_{ij}\,,
\end{equation} 
which invalidates the general MFV 
relation between $b \to s/d$ and $s \to d$ transitions.

Such a large-$\tan\beta$ scenario within the MSSM
was recently studied in Ref.~\cite{Isidori:2006pk}. 
The addditional supersymmetric structure leads to more correlations 
between the observables. 
It  is shown that  for large squark masses and trilinear couplings 
above $1$ TeV this scenario
 explains the present data set naturally, including flavour-conserving 
observables: for example, the significant enhancement of the anomalous 
magnetic moment of the muon $a_\mu$ and the 
large scalar Higgs mass above $115$ GeV, but also the quite  modest 
non-standard contributions to $B^0_s -\bar B^0_s$ mixing and 
to $\bar B \to X_s \gamma$. However, it    
allows for a  large enhancement of  $B \to \mu\mu$ and 
for a significant
suppression  of $B^\pm \to \tau^\pm \nu$, features compatible
with present data. Obviously, the future
measurements of the latter two observables are crucial for the test 
of  this  attractive new-physics scenario.

\begin{figure}[t]
\begin{center}
\includegraphics[width=7.6cm]{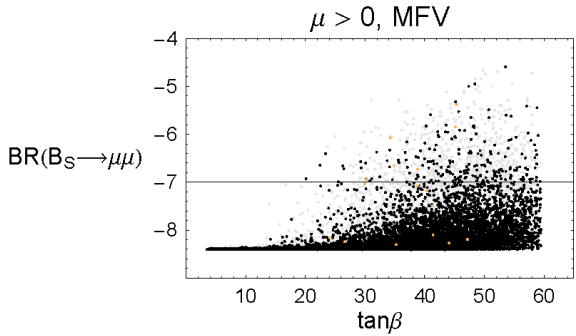}
\vspace{-0.7cm}
\caption{\sf {\cal B}($B_s \to \mu^+ \mu^-$) as a function of $\tan \beta$ in 
a MFV GUT scenario with $\mu > 0$.\label{Enrico}} 
\end{center} 
\end{figure}

\begin{figure}[t]
\begin{center}
\includegraphics[width=7.6cm]{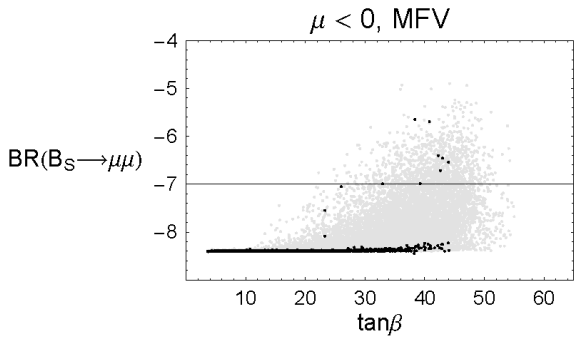}
\vspace{-0.7cm}
\caption{\sf As in Fig.~\ref{Enrico} but $\mu < 0$.\label{EnricoII}} 
\end{center} 
\end{figure}

In Fig.~\ref{Gino}~\cite{Isidori:2006pk}
the constraints of the various $B$ physics observables and of $a_\mu$ 
on the two parameters $\tan \beta$ and the charged Higgs mass $M_H$ 
are shown in an exemplary mode for the Higgs mass parameter $\mu=0.5$, 
the trilinear term $A_U=-1.0$, the average squark mass $M_{\tilde q}=1\,$TeV, $M_1= 0.3\,$TeV, and $M_2= 0.2\,$TeV. 
For the exclusion regions for 
${\cal B}(B_s \to \mu^+\mu^-)$ and ${\cal B}(\bar B \to X_s \gamma)$ 
the  bounds at $90 \%$ c.l.  were taken into account and in the case of 
${\cal B}(B_s \to X_s \gamma)$ the NLL prediction was still used. 
Possible bounds from a future measurement of ${\cal B}( B \to \mu\mu)$ 
and $R_{B \to \tau\nu} = {\cal B}^{\rm exp}(B \to \tau \nu)/{\cal B}^{\rm SM}( B \to \tau \nu)$ are also indicated.

Large-$\tan \beta$ effects in ${\cal B}(B_s \to \mu\mu)$ 
have been analysed, assuming  a supersymmetric minimal flavour violation
structure at the GUT scale~\cite{Lunghi:2006uf}.
The  correlations with other decay modes are  shown in Figs.~\ref{Enrico}
and \ref{EnricoII}, with a mass of the pseudoscalar Higgs of 
$M_A =150\,{\rm GeV}$,  for a Higgs mass parameter $\mu$ negative and 
positive, respectively. The grey points do not survive the constraints  
from $\bar B\to X_s\gamma$ and $a_\mu$. Large enhancement effects are still 
possible  for $\mu > 0$.

Flavour structures beyond the Yukawa  couplings can be introduced into 
the effective  field theory approach with the help of 
 addditional spurions of the flavour group.
Practically, this allows for a different hierarchy between the various 
spurion terms  compared with the one fixed by the structure of the SM.  
Quite recently such a general framework was discussed as next-to-minimal 
flavour  violation scenario (NMFV)~\cite{Feldmann:2006jk}.

In Ref.~\cite{Agashe:2005hk},  a special choice of additional spurions are 
introduced, which modifies only the couplings of the 3rd quark generation.
 This represents  a phenomenologically well-defined class of models; at present those  escape from  the 
most stringent experimental bounds, which are mainly on the first two 
generations.  How minimal flavour violation can be extended  into  the 
framework  of GUT theories  is analysed in Ref.~\cite{Grinstein:2006cg}.

\section{OPPORTUNITIES WITHIN $b \to s$  TRANSITIONS}
Data from $K$ and $B_d$ physics shows that new sources of flavour violation
in \mbox{$s \to d$} and \mbox{$b \to d$} 
are strongly constrained, while 
the possibility of sizable  new contributions to \mbox{$b \to s$} still remains open~\cite{Silvestrini:2005zb,Hurth:2003th}. We also have hints from 
model building; flavour models are not very effective in  constraining 
the \mbox{$b\to s$}  sector~\cite{Masiero:2001cc}. 
Moreover, in supersymmetric grand-unified theories 
the large mixing angle in the neutrino sector relates to  large mixing in 
the right-handed 
$b$--$s$ sector \cite{Moroi:2000tk,Chang:2002mq,Harnik:2002vs}.

Squark  decays are governed  by the same mixing matrices as the
contributions to flavour violating low-energy observables. This allows for 
possible direct correlations between flavour non-diagonal observables 
in $B$ and collider physics. The present bounds on squark mixing, induced by the low-energy data on  $b \to  s$ transitions, still allow for large contributions to flavour violating squark decays at tree level~\cite{Hurth:2003th}.

Among the flavour-changing current processes, the inclusive 
$b \to s \gamma$ and $b \to s \ell^+ \ell^-$
modes  are  still the most prominent. 
The stringent bounds obtained from those  modes
on various non-standard scenarios are a clear example
of the importance of clean FCNC observables in discriminating
new-physics models.
The \mbox{branching} ratio of $\bar B \to X_s \gamma$  has already been measured 
by several independent experiments~\cite{Chen:2001fj,Abe:2001hk,Koppenburg:2004fz,Aubert:2005cu,Aubert:2006gg}, leading  to the 
world average of those five measurements (performed  by
the Heavy Flavour Averaging Group~\cite{unknown:2006bi}) 
for a photon energy cut $E_{\gamma} >
1.6\;{\rm GeV}$:
\begin{eqnarray}
{\cal B}(\bar{B} \to X_s \gamma) =&  \nonumber\\
&\hspace{-2.5cm}= \left(3.55\pm 0.24{\;}^{+0.09}_{-0.10}\pm0.03\right)\times 10^{-4}\,,
\label{hfag}
\end{eqnarray}
where the first error is a combined statistical and systematical  one, 
the second and third   are additional  systematical  errors 
due to the extrapolation  and to the 
$b \to d\gamma$ fraction, respectively. 

After a global effort,  the first theoretical prediction of the 
branching ratio to $O(\alpha_s^2)$ 
has been  recently  presented~\cite{Misiak:2006zs}. 
%The tedious  calculational steps were performed by 
%various groups~\cite{Bieri:2003ue,Misiak:2004ew,Gorbahn:2004my,Gorbahn:2005sa,Melnikov:2005bx,Blokland:2005uk,
%Asatrian:2006ph,Asatrian:2006sm,Czakon:2006notyet,Misiak:2006justnow}. 
For $E_{\gamma} > 1.6\;{\rm GeV}$ the new  prediction 
reads:
\begin{equation} \label{final2}
{\cal B}({\bar B}\to X_s\gamma) = (3.15  \pm 0.23) \times 10^{-4}.
\end{equation}
The overall uncertainty consists of 
non-perturbative (5\%), parametric (3\%), higher-order
(3\%) and $m_c$-interpolation ambiguity (3\%), which have been added 
in quadrature.
Compared with the HFAG average given in Eq.~(\ref{hfag}), the NNLL prediction is $1.2 \sigma$ 
below the experimental data.

The decay $\bar B \to X_s \ell^+\ell^-$ is particularly 
attractive because of  kinematic observables such as 
the invariant dilepton mass spectrum and the forward--backward 
(FB) asymmetry. 
The recently calculated NNLL 
contributions~\cite{Asa1,Adrian2,Adrian1,Asa2,MISIAKBOBETH,Gambinonew}
have significantly 
improved the sensitivity of the inclusive $\bar B \to X_s \ell^+ \ell^-$ decay in  testing extensions of the SM in the sector of flavour 
dynamics; in particular, the value of the dilepton invariant mass
$q^2_0$, for which the differential forward--backward \mbox{asymmetry}
vanishes, is one of the most precise predictions in flavour physics
with a theoretical uncertainty well below $10\%$.

A recent update of the dilepton mass spectrum,  integrated 
over the low dilepton invariant mass region in the muonic case, 
leads to~\cite{Huber:2005ig}
\begin{equation}
{\cal B} (\bar B\to X_s \mu^+\mu^-) =   (1.59\pm 0.11)\times 10^{-6}\,,
\end{equation}
 where the error includes the parametric and perturbative uncertainties only.
The analogous  update of the other NNLL predictions will be presented
in a forthcoming paper~\cite{future}.   

The corresponding rare exclusive decays, such as $B \to K^* \gamma$, 
$B \to K^*  \mu^+ \mu^-$ or also   $B_s \to \phi \mu^+ \mu^-$, 
are experimentally distinguished observables at the forthcoming 
LHCb experiment. 
In contrast to the measurement of the branching ratios, 
measurements of CP, forward-backward, and isospin asymmetries are 
less  sensitive to hadronic uncertainties. Particularly, 
the value of the dilepton invariant mass $q_0^2$, for which the differential 
forward--backward asymmetry
vanishes, can be predicted in quite a clean way. In the QCD
factorization approach, at leading order in $\Lambda_{\rm QCD}/m_b$, the value of 
$q_0^2$ is free from hadronic
uncertainties at order $\alpha_s^0$, a  dependence 
on the soft form factor $\xi_\perp$ and the light-cone wave functions of 
the $B$ and $K^*$ mesons appear at order $\alpha_s^1$. The latter contribution, 
calculated within the  QCD factorization approach,
leads  to a large  shift (see~\cite{Beneke:2001at,Beneke:2004dp,Grinstein:2005ud,Ali:2006ew}).
Nevertheless, there is the well-known issue of power corrections ($1/m_b$) 
within the QCD factorization approach. 
There are also  certain
transversity amplitudes in $B \to K^* \mu^+ \mu^-$, 
which are rather insensitive to 
hadronic uncertainties and in particular highly sensitive to non-standard
chiral  structures of the $b \to s$ current~\cite{Kruger:2005ep}.
For more details on those observables and also on distinguished mixing-induced or direct CP asymmetries 
in $b\to s$ transitions, the reader is referred to a forthcoming 
review~\cite{hurthnew}.

\section{PRESENT STATUS OF THE  SO-CALLED $K\pi$ PUZZLE}
The $B \to K \pi$ modes are well known for  being  sensitive  
to new electroweak penguins beyond the SM~\cite{Fleischer:1997ng,Grossman:1999av}.
The data on CP-averaged $K\pi$ branching ratios can be expressed in 
terms of three ratios:
\begin{eqnarray}
{R}=\frac{\tau_{B^+}}{\tau_{B^0}}
    \frac{{\cal  B}[B^0\to\pi^-K^+]+{\cal B}[\bar B^0\to\pi^+K^-]}
        {{\cal B}[B_d^+\to\pi^+K^0]+{\cal B}[B_d^-\to\pi^-\bar K^0]}\nonumber\\
 {R_n}=\frac12 \frac{{\cal B}[B^0\to\pi^-K^+]+{\cal B}[\bar B^0\to\pi^+K^-]}
        {{\cal B}[B^0\to\pi^0K^0]+{\cal B}[\bar B^0\to\pi^0 \bar K^0]}\nonumber\\
 {R_c}=2\frac{{\cal B}[B_d^+\to\pi^0 K^+]+{\cal B}[B_d^-\to\pi^0 K^-]}
        {{\cal B}[B_d^+\to\pi^+ K^0]+{\cal B}[B_d^-\to\pi^- \bar K^0]}\nonumber
\end{eqnarray}
The actual data presented at ICHEP06 read~\cite{HFAG}

\begin{figure}
\psfig{file=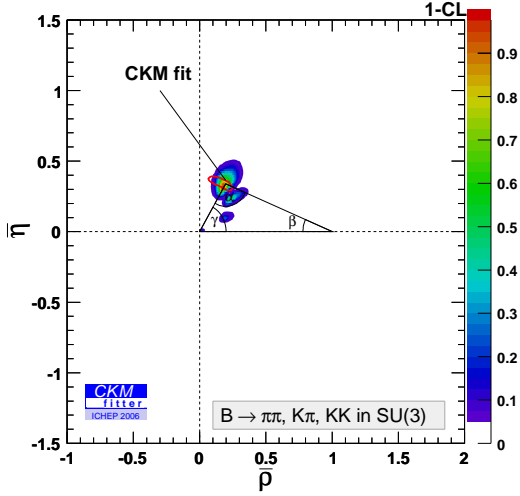,width=7cm} 
\vspace{-1cm}
\caption{\sf Constraint in the $(\bar\rho,\bar\eta)$ plane induced by the 
$\pi\pi,K\pi,K\bar K$ data compared with standard CKM fit.}
\label{charles2}
\end{figure}
\begin{eqnarray}\nonumber
R=0.92^{+0.05}_{-0.05}, R_n=1.00^{+0.07}_{-0.07}, R_c=1.10^{+0.07}_{-0.07},
\end{eqnarray}
which is a  significant change with respect to the pre-ICHEP06 data:
\begin{eqnarray}\nonumber
R=0.84^{+0.06}_{-0.06}, R_n=0.82^{+0.08}_{-0.08}, R_c=1.00^{+0.09}_{-0.09},
\end{eqnarray}
or, with respect to the pre-ICHEP04 data:
\begin{eqnarray}\nonumber
R=0.91^{+0.07}_{-0.07}, R_n=0.76^{+0.10}_{-0.10}, R_c=1.17^{+0.12}_{-0.12}.
\end{eqnarray}
The previous data sets were often called  anomalous 
in view of  the \mbox{approximate} sum rule 
proposed in Refs.~\cite{Lipkin:1998ie,Gronau:1998ep,Matias:2001ch}, 
which leads to the prediction  $R_c = R_n$ and the available 
SM approaches to these data based on QCD factorization techniques and 
on $SU(3)_F$ symmetry assumptions.
The corresponding BBNS predictions,  based on the QCD factorization 
approach~\cite{Beneke:2003zv,BBNSprivate}, are
\begin{eqnarray} \nonumber
R=0.91^{+0.13}_{-0.11}, R_n=1.16^{+0.22}_{-0.19}, R_c=1.15^{+0.19}_{-0.17}.
\end{eqnarray}
Moreover, approximate flavour symmetries 
(isospin or $SU(3)_F$) can also be used 
to relate different decay amplitudes and reduce the number of unknown hadronic
parameters~\cite{Gronau:1990ka,Nir:1991cu}.
In a study~\cite{Buras:2004ub,Buras:2005cv} along 
these lines,  the $B \to \pi\pi$ data were  used to make theoretical
predictions on the $B \to K \pi$ modes. 
This specific  approach   leads to~\cite{Buras:2005cv}
\begin{eqnarray} \nonumber
R=0.96^{+0.02}_{-0.02}, R_n=1.12^{+0.05}_{-0.05}, R_c=1.15^{+0.05}_{-0.05}.
\label{BR}
\end{eqnarray}

The uncertainties 
reflect the experimental uncertainties of the $B \to \pi\pi$ data
only. In the future  the assumptions of  the $SU(3)_F$ 
symmetry can be tested experimentally. 
Because of  the large non-factorizable contributions identified in
the $B \to \pi\pi$ channel, however, large non-factorizable 
$SU(3)_F$- or isospin-violating QCD and QED effects within the SM
cannot be ruled out yet~\cite{Feldmann:2004mg}.

\begin{figure}[t]
\begin{center}
\includegraphics[width=6.4cm,height=6.8cm,angle=90]{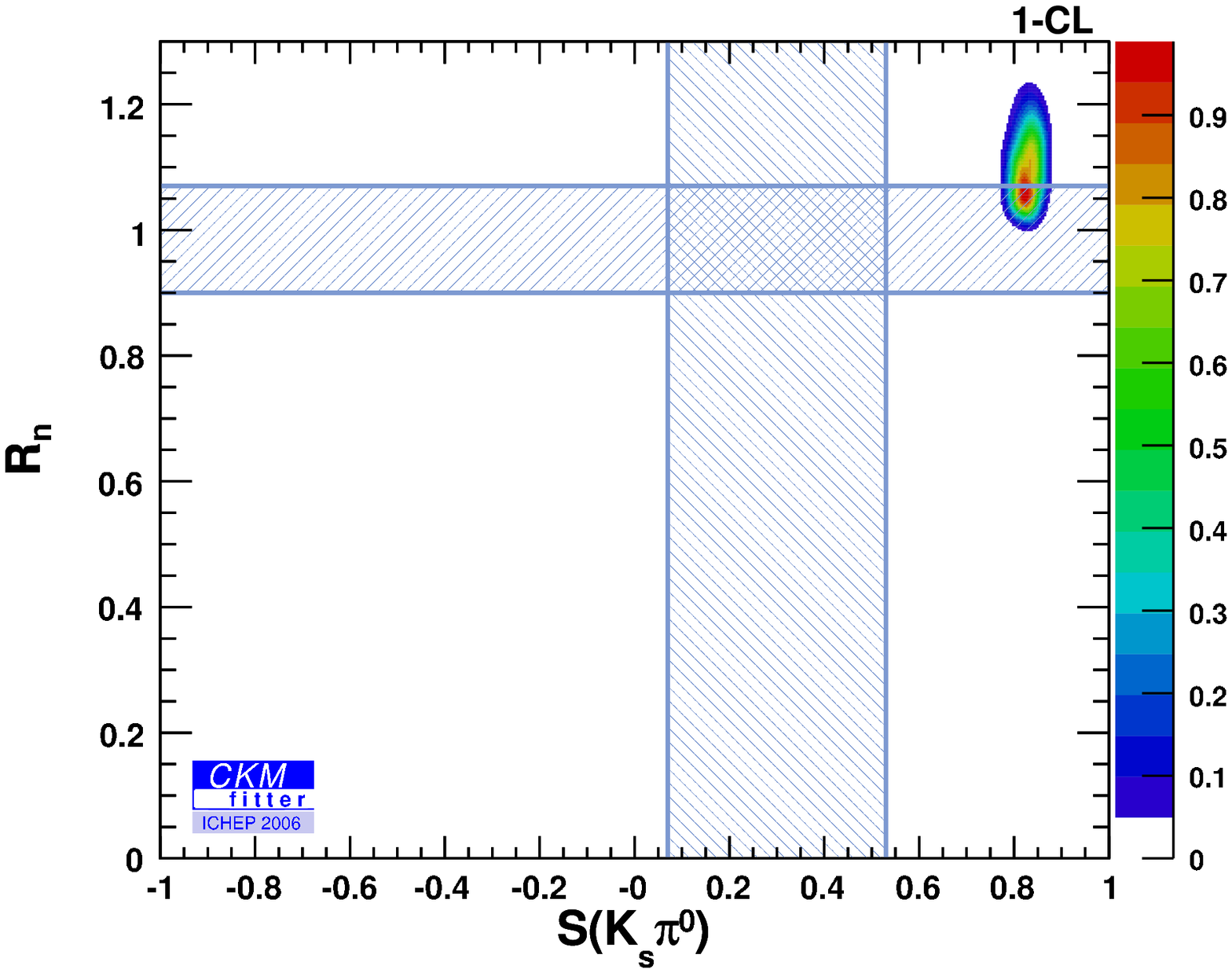}
\vspace{-1cm}
\caption{\sf   Comparison of  direct  measurements of $R_n$ and  $S(K^0_S\pi^0)$
($1\sigma$ band) with indirect fit ($2\sigma$ contour).   \label{charles1}} 
\end{center} 
\end{figure}

Nevertheless, the new data set~\cite{HFAG} significantly moved into the 
ballpark of the SM estimates. In the previous analyses  
the radiative corrections to charged particles in the final state 
were  not taken into account,  as 
was emphasized in the past (see for example Ref.~\cite{Hurth:2005wg}). 
 These corrections, worked out in Ref.~\cite{Baracchini:2005wp},  
are now properly included in the analysis of both 
experiments and are partially responsible 
for the shifts in the central values in the $K\pi$ data.

\begin{figure}[t]
\begin{center}
\includegraphics[width=7cm,height=7cm,angle=0]{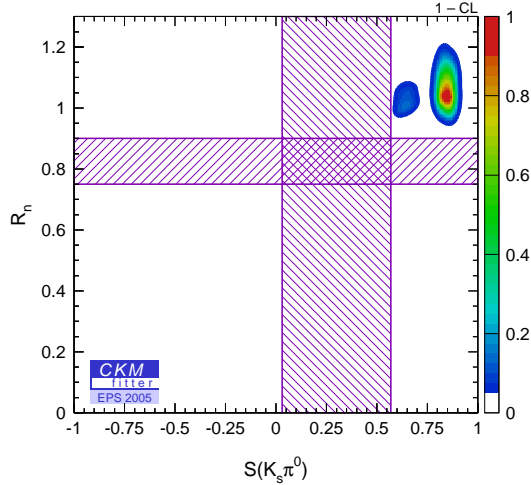}
\vspace{-1cm}
\caption{\sf  As Fig.~\ref{charles1}, but  based on pre-ICHEP06 data. \label{charles1old}} 
\end{center} 
\end{figure}

A  new, more complete $SU(3)_F$ analysis of 
the CKM fitter group is now accessible~\cite{charlesref},
in which all  available $\pi\pi,K\pi,K\bar K$ modes are  
\mbox{included,} \mbox{so-called}  annihilation/exchange topologies and  
factorizable $SU(3)_F$ breaking are taken into account. 
As shown  in Fig.~\ref{charles2}~\cite{charlesref}
the constraint in the $(\bar\rho,\bar\eta)$ plane induced by these data
implies that the compatibility with  the $SU(3)$ and 
SM hypothesis is very good (the so-called pValue of that 
SM analysis is of order  $30-40\%$). 
But  the $\chi^2_{\rm min}$ is not always the best measure of the 
compatibility of the data with the theory. Among the main 
contributions to the $\chi^2$  there  are 
the ratio $R_n$  
and the CP asymmetry $S(K^0_S\pi^0)$, which 
are all very sensitive to new electroweak penguins.    
After removing them from the global fit,  
Fig.~\ref{charles1}~\cite{charlesref}  
shows the  comparison 
of the indirect fit ($2\sigma$ contour),  with $\bar \rho, \bar \eta$ 
from the CKM fit and  all  other available modes, with the
direct measurements  ($1\sigma$ band) using the new  
data. This can be  compared with the analogous plot based on the 
pre-ICHEP2006 data, see Fig.~\ref{charles1old}.   
While the indirect prediction for $R_n$ is now in good agreement with the 
direct measurement, there is still a small `discrepancy' 
in the case of the observable $S(K^0_S\pi^0)$.

Future data from the $B$ factories and LHCb  will clarify the situation 
completely. There will be  up to 38 measured observables  depending on the 
same 13+2 theoretical parameters. This will allow for the study of $SU(3)$ breaking  and new-physics effects.

\section*{ACKNOWLEDGEMENTS}
We thank Gino Isidori, Thorsten Feldmann, and Mikolaj Misiak 
for useful  discussions and a careful reading of the manuscript.

\end{document}